\documentclass[11pt]{article}
\usepackage{moriond}

\def\be{\begin{equation}}
\def\ee{\end{equation}}
\def\bea{\begin{eqnarray}}
\def\eea{\end{eqnarray}}

\usepackage{fancyhdr}
\usepackage{amsmath}
\usepackage{latexsym}
\usepackage{amssymb}
\usepackage{epsf}
\usepackage{graphicx}
\usepackage{slashed}

\newcommand{\Photo}{\includegraphics[height=35mm]{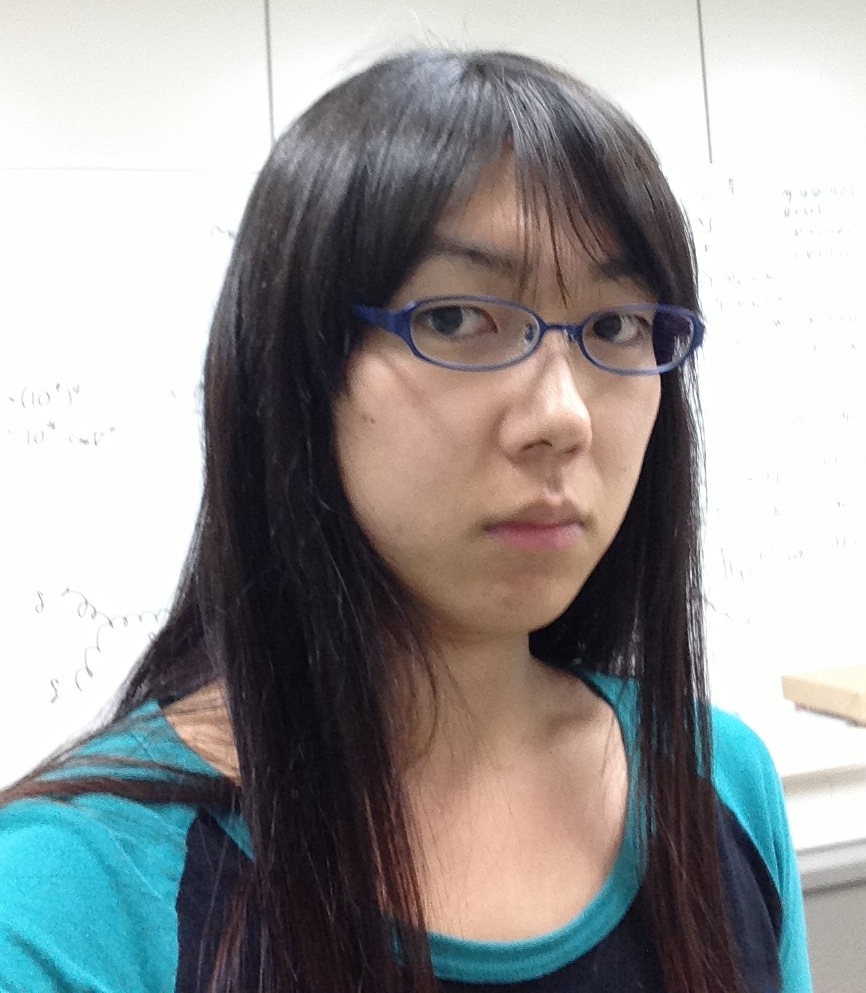}}

\begin{document}
\pagestyle{plain}

\vspace*{4cm}
\title{Constraint on Universal Extra Dimensions from scalar boson searches}

\author{
Takuya Kakuda$^1$\footnote{This talk is given by T. Kakuda in Rencontres de Moriond EW 2013.},
\ 
Kenji Nishiwaki$^2$,
\ 
Kin-ya Oda$^3$,
\ and
Ryoutaro Watanabe$^4$}

\address{\it $^1$ Graduate school of science and technology, Niigata University, Niigata 950-2181, Japan \\
\it $^2$ Regional Centre for Accelerator-based Particle Physics,\\
\it Harish-Chandra Research Institute, Allahabad 211 019, India\\
\it $^3$ Department of Physics, Osaka University, Osaka 560-0043, Japan\\
\it $^4$ Theory Group, KEK, Tsukuba, Ibaraki 305-0801, Japan}

\maketitle\abstract{
We show the bounds on five- and six-dimensional Universal Extra Dimension models from the result of the Higgs boson searches at the Large Hadron Collider and electroweak precision measurement. 
The latest data released by the ATLAS and the CMS gives the lower bounds on Kaluza-Klein scale which are from 650 GeV to 1350 GeV depending on models 
from Higgs to diboson/diphoton decay signal.
The Higgs production cross section can be enhanced by factor 1.5 in {crude} estimation, diphoton decay signal is suppressed about 10$\%$.
Electroweak precision measurement also gives the lower bounds as {from} 700 GeV to 1500 GeV.
This is a proceedings of the conference ``Rencontres de Moriond EW 2013".
}

\section{Introduction}
The ATLAS and CMS experiments reported their recent results on signal strengths of the Higgs like boson 
for its decay into diphoton ($\gamma\gamma$) and diboson ($ZZ$ and $WW$)~\cite{ATLAS:2013gamma,ATLAS:2013Z,ATLAS:2013W,CMS:2013gamma,CMS:2013Z,CMS:2013W}. 
{The signal strengths} of $H \to \gamma\gamma$, $ZZ$ and $WW$ turn out to be $1.65\pm0.24^{+0.25}_{-0.18}$, $1.7^{+0.5}_{-0.4}$ and $1.01\pm0.31$ at the ATLAS experiment, 
while $0.78\pm0.27$ (MVA based), $0.91^{+0.30}_{-0.24}$ and $0.71\pm0.37$ (cut based) at the CMS experiment. 
These results are consistent with the SM but there still is a room for a new physics effect in these processes. 
In this work, we put bounds on universal extra dimension (UED) models.

The UED is a candidate of new physics, in which all the SM particles propagate in extra compactified spacial dimensions. 
The five-dimensional minimal UED (mUED) model without tree-level brane-localized term as a minimal extension of the SM, which is constructed on $S^1/Z_2$~\cite{Appelquist:2000nn}, has been well studied.
Six-dimensional UED models with various two-dimensional compactified spaces are also considered. 
We investigate the 6D UED models based on two torus, $T^2/Z_2$~\cite{Appelquist:2000nn}, $T^2/Z_4$~\cite{Dobrescu:2004zi,Burdman:2005sr}, $T^2/(Z_2 \times Z'_2)$~\cite{Mohapatra:2002ug}, on two sphere $S^2/Z_2$~\cite{Maru:2009wu} and $S^2$ {with Stueckelbarg field}, and on the non-orientable manifolds, namely the real projective plane $RP^2$~\cite{Cacciapaglia:2009pa} and the projective sphere (PS)~\cite{Dohi:2010vc}, 
by putting bounds on the Kaluza-Klein (KK) scale from the results of the Higgs signal search and the electroweak precision measurements.
For details of these models, see for example Refs.~\cite{Nishiwaki:2011gm,Nishiwaki:2011gk}.

For bounds on the UED models from the electroweak precision measurements, we use the $S$ and $T$ parameters.
The recent constraints on the $S$ and $T$ parameters are given in Ref.~\cite{Baak:2012kk}. 
For the bounds from the Higgs signal search, we use the recent results obtained in Ref.~\cite{ATLAS:2013gamma,ATLAS:2013Z,ATLAS:2013W,CMS:2013gamma,CMS:2013Z,CMS:2013W} for each decay process. 
In order to calculate these quantities in the UED models, we need to know an ultraviolet (UV) cutoff scale in a view point of four-dimensional effective theory. 
To search for the highest possible UV cutoff scale, we have evaluated the vacuum stability bound on the Higgs potential by solving renormalization group equation (RGE). 


\section{RGE and vacuum stability bound in UED}

Let us review how to compute RGE in a theory with compactified extra dimension(s).
We adopt the bottom-up approach {in} Refs.~\cite{Bhattacharyya:2006ym,Ohlsson:2012hi},
{where we take} into account a 
contribution of a massive particle to the beta functions when the increasing scale {$\mu$} passes its mass. 
In the case of the UED, after KK decomposition, the corresponding 4D effective theory
contains not only the SM fields, but also their KK partners. 
Following this prescription, we get the beta function of coupling constant $c$ as
\begin{equation}
\beta_c = \beta_c^{\text{(SM)}} + \sum_{s:\,\text{massive states}} \theta(\mu - M_s) \Big( N_s
\beta^{\text{(NP)}}_{s,c} \Big),
\label{Eq:betafunction}
\end{equation}
where $\beta_c^{\text{(SM)}}$ and {$\beta_{s,c}^{\text{(NP)}}$} are the contributions from the SM particles and 
from the new massive ones with mass $M_s$, respectively, and $N_s$ is the number of degenerated states in the KK state $s$.
The vacuum stability bound can be evaluated by solving RGE {as the point where the running Higgs self-coupling $\lambda(\mu)$ turns out to be zero}. 
{Table}~\ref{table:maximalcutoff} shows the vacuum stability bound {$\Lambda_{\text{max}}$} for each {model} in the case of {the KK scale $M_{\text{KK}} = 1\,\text{TeV}$}.
Note that {the values of $\Lambda_{\text{max}}$ are almost universal within the case of $M_{\text{KK}} \sim$ a few TeV}.
{The details are found} in~\cite{KNOOW:UED2013}.
{In the following analyses, we employ the numbers in Table~\ref{table:maximalcutoff}} as the UV cutoff scale in {four-dimensional} effective theory.

\begin{table}[t]
\begin{center}
\begin{tabular}{|c||c|c|c|c|c|c|c|c|}
\hline
model &
mUED & 
$T^2/Z_2 $ & 
$T^2/{(Z_2 \times Z'_2)}$ & 
$T^2/Z_4$ & 
$S^2 $ & 
$S^2/Z_2$ & 
$RP^2$ & 
{PS} \\ \hline \hline
$\tilde{\Lambda}_{\text{max}}$ &
5.0 & 2.5 & 2.9 & 3.4 & 2.3 & 3.2 & 2.0 & 1.9 \\ \hline
\end{tabular}
\caption{Upper bounds on cutoff scale $\Lambda_{\text{max}} = \tilde{\Lambda}_{\text{max}} M_{\text{KK}}$ with $M_{\text{KK}} = 1\,\text{TeV}$.}
\label{table:maximalcutoff}
\end{center}
\end{table}

\section{Higgs signals at the Large Hadron Collider}

The Higgs signal at the LHC can be divided into two parts, production and decay processes.
Higgs production at the LHC mainly comes from gluon fusion through the top loop.
On the other hand, Higgs to diphoton and digluon decays are also induced as loop processes that are mainly constructed by top and W boson loops.
KK tower in the UED models affects such loop processes. 
%
We compute a signal strength for the gluon fusion production channel in the UED models:
\begin{equation}
\mu_{gg \to H \to X}=  
  \frac{\sigma ^\text{UED}_{gg\rightarrow H \rightarrow X}}{\sigma ^\text{SM}_{gg\rightarrow H \rightarrow X}} 
  \simeq \frac{\Gamma ^\text{UED}_{H \rightarrow gg} \Gamma ^\text{UED}_{H\rightarrow X} /\Gamma ^\text{UED}_{H}}
          {\Gamma ^\text{SM}_{H \rightarrow gg} \Gamma ^\text{SM}_{H\rightarrow X} / \Gamma ^\text{SM}_{H}} \label{signal-st-2gamma},
\end{equation}
where $X=\gamma \gamma,ZZ,WW,$ etc, {$\Gamma^{\text{UED/SM}}_{H}$} is total decay width of the Higgs in UED/SM case. 
For {each model} we analyzed, $\Gamma ^\text{UED}_{H \rightarrow gg}$ is enhanced by KK top loops comparing with the SM.
For the final states $X=ZZ/WW$, we can approximate as $\Gamma ^\text{UED}_{H\rightarrow ZZ/WW} \sim  \Gamma ^\text{SM}_{H\rightarrow ZZ/ WW}$ 
because Higgs decays into $ZZ/WW$ boson pair at the tree level, and hence KK loop contributions are negligible.
For the final state $X=\gamma \gamma$, the sum of KK towers supress $ \Gamma ^\text{UED}_{H\rightarrow \gamma \gamma}$.
The reason of the suppression is as follows. 
Each KK fermion mode is vectorlike, and hence has twice the degrees of freedom compared to its zero mode. 
Therefore their negative contributions to decay rate become larger than the positive ones coming from the KK $W$ loops~\cite{Nishiwaki:2011gm,Nishiwaki:2011vi}.
{
If we consider the {KK scale as $1\,\text{TeV}$}, the signal strength of $H \to \gamma \gamma$ is enhanced by a factor 1.5 from that of the SM, 
while the diphoton decay rate is suppressed by 10\%. Thus, $\mu_{gg \to H \to \gamma\gamma}$ becomes about 1.35 times larger than that in the SM.
The details of the enhancement and suppression are given in Ref.~\cite{Nishiwaki:2011gm,Nishiwaki:2011vi,KNOOW:UED2013}.}

\begin{figure}[t]
\vspace{-6mm}
\begin{tabular}{cc}

\begin{minipage}{0.5\hsize}
\begin{center}
   \includegraphics[clip,width=70mm]{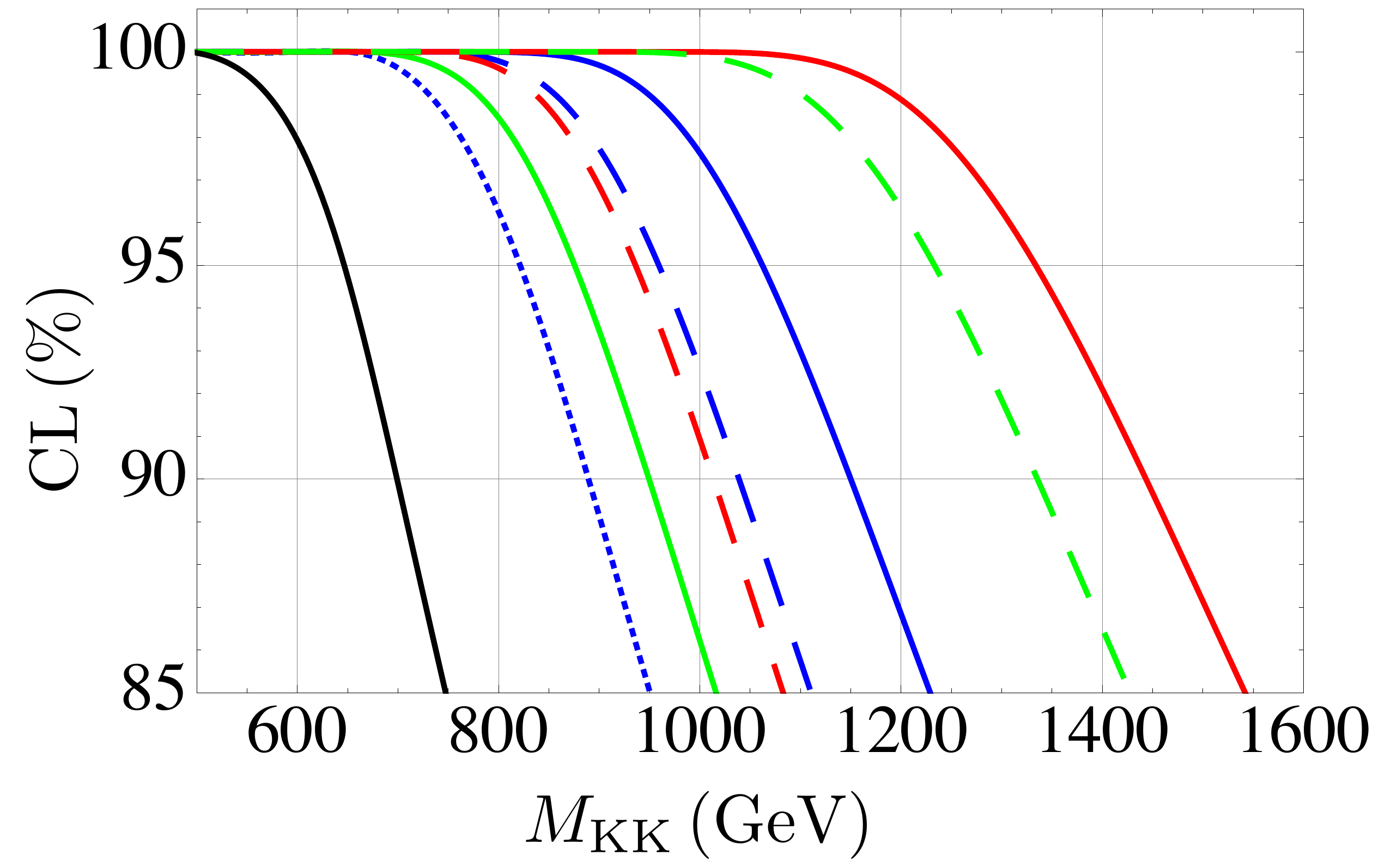}
\label{Fig:Higgs}
\end{center}
\end{minipage}

\begin{minipage}{0.5\hsize}
\begin{center}
   \includegraphics[clip,width=70mm]{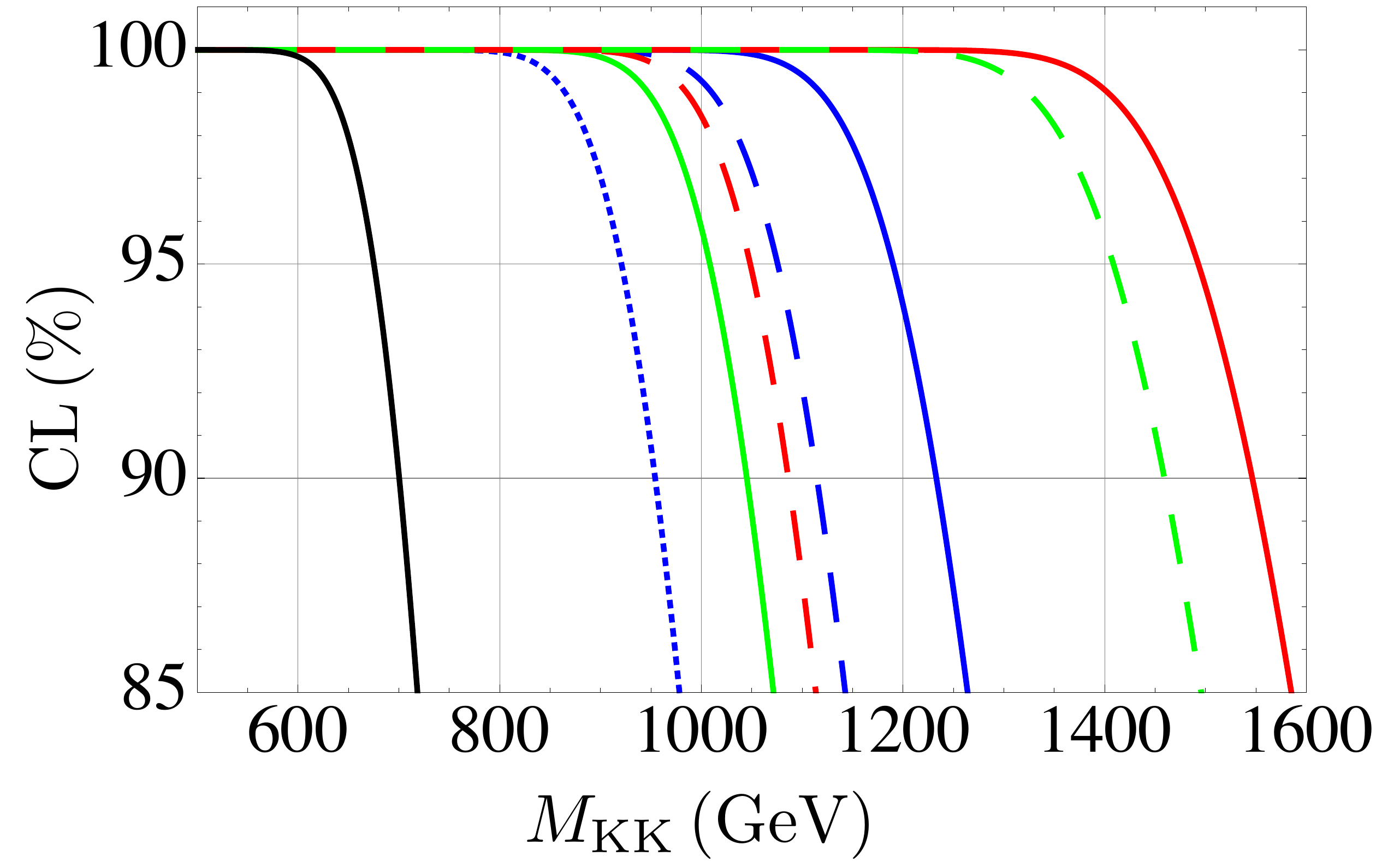}
\label{Fig:STresuls}
\end{center}
\end{minipage}

\end{tabular}
\caption{An exclusion CL of all UED models as a function of the KK scale
{$M_{\text{KK}}$} by use of all the ATLAS and CMS results of $H \rightarrow \gamma \gamma,WW,ZZ$ (left), 
and recent {$S$, $T$} result (right){,} where the blue line,
dashed line, and dotted line show the results in $T^2/Z_2, T^2/Z_2 \times Z_2'
 $, and $T^2/Z_4$ models; the red line and dashed line show that in $S^2$ and $S^2/Z_2$ models; the green line and dashed line 
show that in $RP^2$ and $PS$ models; and the black line indicate that in mUED model.}
\label{Fig:HiggsST}
\end{figure}

As shown above, the UED models give different production cross section in the gluon fusion (GF).
On the other hand, the other productions: the vector boson fusion (VBF),
the Higgs-strahlung (VH), and the associated production with a $t \bar{t}$ pair (ttH) are the same as in the SM.
The ATLAS and CMS have reported on the proportions of these production channels for {each event category} of $H \rightarrow \gamma \gamma ,ZZ$ and $WW$~~\cite{ATLAS:2013gamma,ATLAS:2013Z,ATLAS:2013W,CMS:2013gamma,CMS:2013Z,CMS:2013W}.
We take these contributions into account. The details are given in~\cite{KNOOW:UED2013}.
The left of Figure \ref{Fig:HiggsST} shows the bounds on the KK scale from all the ATLAS and CMS results of $H \to \gamma \gamma,WW,ZZ$ channels.


\section{Electroweak precision test}
A measurement related to electroweak sector can be used to obtain indirect bounds on phenomenological models. 
The $S$ and $T$ parameters 
are very useful quantities for such a purpose. 
These parameters are represented by two point functions of the gauge bosons.
Several measurable quantities are represented as functions of the $S$ and $T$ parameters,
and from the global fit to the experimental results, the values of $S,T$ are estimated as~\cite{Baak:2012kk} 
\begin{equation}
S|^\text{exp}_{U=0} = 0.05 \pm 0.09,\quad
T|^\text{exp}_{U=0} = 0.08 \pm 0.07,
\label{ST_experimental}
\end{equation}
with {$126\,\text{GeV}$ reference Higgs mass and} its correlation being $+0.91$, assuming that the $U$ parameter is zero. 
In an operator-analysis point of view, the $U$ parameter is represented as a coefficient of a higher dimensional operator involving the Higgs doublet than those for $S$ and $T$ in the UED models, 
and hence we ignore the effect in our analysis.
%
The KK top and KK Higgs contributions to $S$ and $T$ {in the mUED} are already estimated in~{\cite{Appelquist:2002wb,Baak:2011ze}}.
Our analysis newly take into acount the effect of KK gauge bosons.
The detailed forms of $S,T$ in UED models are found in Ref.~\cite{KNOOW:UED2013}.
The right of Figure~\ref{Fig:HiggsST} shows the bounds on the KK scales from the fit to the results in Eq.~(\ref{ST_experimental}). 


\section{Summary}
We have estimated the two types of bounds on the KK scales in 5D and 6D UED models from the Higgs boson searches at the LHC 
and from the electroweak precision data via the $S$ and $T$ parameters.
Due to the contributions of the KK loops including KK top and KK gauge bosons, the Higgs decay ratio and production cross section are modified.
The KK excited states of the massive SM particles (top quark, Higgs boson and gauge boson) alter the $S$ and $T$ parameters. 
Comparing the bounds from the Higgs signal search with those from the electroweak measurements, 
we find that the latter bounds are slightly severer than the former ones in the UED models for now.
However, in future the Higgs signal search at the LHC will put more strong constraints on the KK scales in the UED models.

\subsection*{Acknowledgement}
{We thank Tomohiro Abe for useful comments on oblique corrections and 
Swarup Kumar Majee for discussions at the early stages of this work.
K.N.\ is grateful for valuable discussions with Joydeep Chakrabortty and Daisuke Harada, and for fruitful conversations with Anindya Datta and Sreerup Raychaudhuri.
K.N.\ is partially supported by funding available from the Department of 
Atomic Energy, Government of India for the Regional Centre for Accelerator-based
Particle Physics (RECAPP), Harish-Chandra Research Institute.
The work of K.O. (R.W.) is in part supported by the Grands-in-Aid for Scientific Research No.~23104009, No.~20244028, and No.~23740192 (No.~248920).}


\section*{References}

\bibliography{referenceMoriond}

%
%
%
%

\end{document}